\documentstyle[12pt]{article}
\begin{document}
\huge
\begin{center}
\Large{ \bf BLACK HOLE EMISSION IN STRING THEORY \\  AND THE STRING PHASE
OF BLACK HOLES}\\ 
\vspace{1cm}
{\bf N. SANCHEZ} \footnote{Email: Norma.Sanchez@obspm.fr}

\vspace{1cm}
{Observatoire de Paris, Demirm/Lerma, Laboratoire Associ\'e au CNRS, Observatoire de
Paris et Ecole Normale Sup\'erieure,  \\
61 Avenue de l'Observatoire, 75014 Paris,
France. \\}
\begin{abstract}
\normalsize
The quantum string emission by Black Holes is computed in the
framework of the `string analogue 
model' (or thermodynamical approach), which is well suited to combine
quantum field theory (QFT) and string theory in curved 
backgrounds (particulary here, as black holes and strings posses
intrinsic thermal features and temperatures).  

String theory properly describes black-hole evaporation. The
black-hole temperature is the Hawking temperature in
the semiclassical (QFT) regime and  becomes the intrinsic string
temperature, $T_s$, in the quantum (last stage) string regime.

The QFT-Hawking temperature $T_H$ is upper bounded by the string
temperature $T_S$ in  
the black hole background. The black hole emission spectrum is an
incomplete gamma function of  
($T_H - T_S$). For $T_H \ll T_S$, it yields the QFT-Hawking
emission. For $T_H \to T_S$, it shows 
highly massive string states dominate the emission and undergo a
typical string phase transition to a  
{\it microscopic} `minimal' black hole of mass $M_{\min}$ or radius
$r_{\min}$ (inversely proportional  
to $T_S$) and string temperature $T_S$. \\
The semiclassical QFT black hole (of mass $M$ and temperature $T_H$)
and the string black hole (of mass 
$M_{\min}$ and temperature $T_S$) are mapped one into another by a
`Dual' transform which links  
classical/QFT and quantum string regimes. \\
The string back reaction effect (selfconsistent black hole solution of
the semiclassical Einstein 
equations with mass $M_+$ (radius $r_+$) and temperature $T_+$)
 is computed. Both, the QFT and string
black hole regimes are well defined and bounded: $r_{\min} \leq r_+
\leq r_S \: , \: M_{\min} \leq 
M_+ \leq M \: , \: T_H \leq T_+ \leq T_S$. \\
The string ``minimal'' black hole has a life time $\tau_{\min} \simeq
\frac{k_B c}{G \hbar} \: T^{-3}_S$. 
\end{abstract}
\end{center}
\normalsize

\newpage

\section{INTRODUCTION AND RESULTS}

In the context of Quantum Field Theory in curved spacetime, Black
Holes have an intrinsic Hawking  temperature Ref. [1] given by
$$
T_H = \: \frac{\hbar c}{4 \pi k_B} \: \frac{(D-3)}{r_S} \qquad
\qquad \qquad \qquad r_S \equiv L_{cl} 
$$

\noindent
$r_S$ being the Schwarzchild's radius (classical length $L_{cl}$). \\

In the context of Quantum String Theory in curved spacetime, quantum
strings in black hole spacetimes have an intrinsic temperature given by

$$T_S = \frac{\hbar c}{4 \pi k_B} \: \frac{(D-3)}{L_q} \: , \: L_q = \frac{b L_S (D-3)}
{4 \pi} \: , \: L_S \equiv \sqrt{\frac{\hbar \alpha^{\prime}}{c}} \: \: ,$$

\noindent
which is the same as  the string temperature in flat spacetime (See
Ref. [2] and Section 3 in this  
paper).  

The QFT-Hawking temperature $T_H$ is a measure of the Compton length
of the Black Hole, and thus, of 
its ``quantum size'', or quantum property in the semiclassical-QFT
regime. The Compton length of a 
quantum string is a direct measure of its size $L_q$. The string
temperature $T_S$ is a measure of the  
string mass, and thus inversely proportional to $L_q$. 

The ${\cal R}$ or ``Dual'' transform over a length introduced in
Ref. [3] is given by:

$$\begin{array}{ccc}  \tilde{L}_{cl} &= {\cal R} L_{cl} &= L_q \\
\tilde{L}_q &= {\cal R} L_q &= L_{cl} \end{array}$$

\medskip
\noindent
Under the ${\cal R}$-operation:

$$\tilde{T}_H = T_S$$

\noindent
and

$$\tilde{T}_S = T_H$$

The QFT-Hawking temperature and the string temperature in the black hole background are 
${\cal R}$-Dual of each other. This is valid in all spacetime dimensions $D$, and is a generic
feature of QFT and String theory in curved backgrounds, as we have shown this relation for the
respectives QFT-Hawking temperature and string temperature in de Sitter space Ref. [3]. In fact, the
${\cal R}$-transform maps QFT and string domains or regimes.

In this paper, we investigate the issue of Hawking radiation and the back reaction effect on the black
hole in the context of String Theory. In principle, this question should be properly addressed in the
context of String Field Theory. On the lack of a tractable framework for it, we work here in the
framework of the string analogue model (or thermodynamical approach). This is a suitable approach for
cosmology and black holes in order to combine QFT and string study and to go further in the 
understanding of quantum gravity effects. The thermodynamical approach is particularly appropriated 
and natural for black holes, as Hawking radiation and the string gas [4,5] posses intrinsic thermal
features and temperatures.

In this approach, the string is a collection of fields $\Phi_n$ coupled to the curved background, 
and whose masses $m_n$ are given by the degenerate string mass spectrum in the curved space considered.
Each field $\Phi_n$ appears as many times the degeneracy of the mass level $\rho (m)$. (Althrough the
fields $\Phi_n$ do not interact among themselves, they do with the black hole background).

In black hole spacetimes, the mass spectrum of strings is the same as in flat spacetime Ref. [2], 
therefore the higher masses string  spectrum satisfies

$$\rho (m) = \left( \sqrt{\frac{\alpha^\prime c}{\hbar}} \: m \right)^{-a} \: 
e^{b\sqrt{\frac{\alpha^\prime c}{\hbar}} \: m} $$

\medskip
\noindent
($a$ and $b$ being constants, depending on the model, and on the number of space dimensions).

We consider the canonical partition function ($\ln Z$) for the higher excited quantum string states of
open strings (which may be or may be not supersymmetric) in the asymptotic (flat) black hole region. The gas of
strings is at thermal equilibrium with the black hole at the Hawking temperature $T_H$, it follows that
the canonical partition function, [Eq. (9)] is well defined for Hawking temperatures satisfying the
condition

$$T_H < T_S$$

\medskip
\noindent
$T_S$ represents a maximal or critical value temperature. This limit implies a minimum horizon radius

$$r_{\min} = \frac{b (D-3)}{4 \pi} \: L_S$$

\medskip
\noindent
and a minimal mass for the black hole ($BH$).

$$\begin{array}{l} M_{\min} = \frac{c^2 (D-2)}{16 \pi G} \: A_{D-2} \: r_{\min}^{D-3}  
\\
\left( M_{\min} (D = 4) = \frac{b}{8 \pi G} \: \sqrt{\hbar c^3 \alpha^{\prime}} \right)
\end{array}$$

We compute the thermal quantum string emission of very massive particles by a $D$ dimensional 
Schwarzschild $BH$. This highly massive emission, corresponding to the higher states of the string
mass spectrum, is naturally expected in the last stages of $BH$ evaporation.

In the context of QFT, $BH$ emit particles with a Planckian (thermal)
spectrum at temperature $T_H$. 
The quantum $BH$ emission is related to the classical absorption cross
section through the Hawking formula Ref. [1]:
$$
\sigma_q (k,D) = \frac{\sigma_A (k,D)}{e^{E(k)/k_B T_H} - 1} 
$$
The classical total absorption spectrum $\sigma_A (k,D)$ Ref. [6] is
entirely oscillatory as a function 
of the energy. This is exclusive to the black hole (other absorptive
bodies do not show this property).  

In the context of the string analogue model, the quantum emission by
the $BH$ is given by

$$\sigma_{\rm string} (D) = \: \sqrt{\frac{\alpha^\prime c}{\hbar}} \: \int^\infty_{m_0} \: 
\sigma_q (m,D) \: \rho (m) dm $$

\medskip
\noindent
$\sigma_q (m,D)$ being the quantum emission for an individual quantum field with mass $m$ in the
string mass spectrum. $m_0$ is the lowest mass from which the asymptotic expression for $\rho (m)$ is
still valid.

We find $\sigma_{\rm string} (D)$ as given by [Eq. (34)] (open strings). It consists of two terms:
the first term is characteristic of a quantum thermal string regime, dominant for $T_H$ close to
$T_S$;  the second term, in terms of the exponential-integral function $E_i$ is dominant for $T_H \ll
T_S$ from which the QFT Hawking radiation is recovered. \\
For $T_H \ll T_S$ (semiclassical QFT regime):

$$\sigma^{({\rm open})}_{\rm string} \simeq B(D) \beta_H^{\frac{(D-5)}{2}} \: 
\beta_S^{-\frac{(D-3)}{2}} \: e^{- \beta_H m_0 c^2} \: , \: \beta_H \equiv (T_H k_B)^{-1} $$

For $T_H \to T_S$ (quantum string regime):

$$\sigma_{\rm string} \simeq B(D) \: \frac{1}{( \beta_H - \beta_S )} \: , \: \beta_S \equiv (T_S 
k_B)^{-1}$$

\medskip
\noindent
$B(D)$ is a precise computed coefficient [Eq. (34.a)].

The computed $\sigma_{\rm string} (D)$ shows the following: At the first stages, the $BH$ emission 
is in the lighter particle masses at the Hawking temperature $T_H$ as described by the semiclassical 
QFT regime (second term in [Eq. (34)]. As evaporation proceeds, the temperature increases, the $BH$
radiates the higher massive particles in the string regime (as described by the first term of 
[Eq. (34)]). For $T_H \to T_S$, the $BH$ enters its quantum string regime $r_S \to r_{\min}$, $M \to
M_{min}$.  

That is, ``the $BH$ becomes a string'', in fact it is more than that, as [Eq. (34)] accounts for the 
back reaction effect too: The first term is characteristic of a Hagedorn's type singularity Ref. [5],
and the partition function here has the same behaviour as this term. Its meaning is the following: At
the late stages, the emitted $BH$ radiation (highly massive string gas) dominates and undergoes a 
Carlitz's type phase transition Ref. [5] at the temperature $T_S$ into a condensed finite energy state.
Here such a state (almost all the energy concentrated in one object) is a {\it microscopic} (or
``minimal'') $BH$ of size $r_{\min}$, (mass $M_{\min}$) and temperature $T_S$.

The last stage of the $BH$ radiation, properly taken into account by string theory, makes such a phase
transition possible. Here the $T_S$ scale is in the Planck energy range and the transition is to a 
state of string size $L_S$. The precise detailed description of such phase transition and such final 
state deserve investigation.   

A phase transition of this kind has been considered in Ref. [7]. Our results here supports and give
a precise picture to some issues of $BH$ evaporation discussed there in terms of purely 
thermodynamical considerations.

We also describe the (perturbative) back reaction effect in the framework of the semiclassical Einstein 
equations ($c$-number gravity coupled to quantum string matter) with the v.e.v. of the energy 
momentum tensor of the quantum string emission as a source. In the context of the analogue model, such
stress tensor v.e.v. is given by:

$$\langle \tau_\mu^\nu (r) \rangle = \: \frac{\int^\infty_{m_0} \: \langle T_\mu^\nu (r,m) \rangle 
\sigma_q (m,D) \rho (m) dm}{\int^\infty_{m_0} \: \sigma_q (m,D) \rho (m) dm}$$

\medskip
\noindent
Where $<T_\mu^\nu (r,m)>$ is the v.e.v. of the QFT stress tensor of individual quantum fields of mass
$m$ in the higher excited string spectrum. The solution to the semiclassical Einstein equations is 
given by ([Eq. (53)], [Eq. (56)], [Eq. (60)]) ($D=4$):

$$r_+ = r_S \left( 1 - \: \frac{4}{21}  \: \frac{{\cal A}}{r^6_S} \right)  $$

$$M_+ = M \left( 1 - \: \frac{4}{21}  \: \frac{{\cal A}}{r^6_S} \right)  $$

$$T_+ = T_H \left( 1 + \: \frac{1}{3}  \: \frac{{\cal A}}{r^6_S} \right)  $$

\medskip
\noindent
The string form factor ${\cal A}$ is given by [Eq. (62)], it is finite and positive. For $T_H \ll T_S$, 
the back reaction effect in the QFT-Hawking regime is consistently recovered. Algebraic terms in
($T_H - T_S$) are enterely stringly. In both cases, 
the relevant ratio ${\cal A} / r^6_S$ entering in 
the solution ($r_+ , M_+ , T_+$) is negligible. 
It is illustrative to show it in the two
opposite regimes:

$$\begin{array}{cccc}
\left( \frac{{\cal A}}{r^6_S} \right)^{\rm open / closed} \: 
& \simeq & \frac{1}{80640 \pi} \:
\left( \frac{M_{PL}}{M} \right)^4 \: \left( \frac{M_{PL}}{m_0} 
\right)^2 \: \ll 1  \\
& T_H \ll T_S & \\
& & \\
\left( \frac{{\cal A}}{r^6_{\min}} \right)^{\rm closed} \: & \simeq & \frac{16}{735 b} \:
\left( \frac{\pi}{b} \right)^3 \: \left( \frac{M_S}{M_{PL}} \right)^2 \: \left( \frac{M_S}{m_0}
\right)^2 \: \ll 1  \\
& T_H \to T_S & \\
\end{array}$$

\medskip
\noindent
$M_{PL}$ being the Planck mass and $M_S = \frac{\hbar}{c L_S}$.

The string back reaction solution shows that the $BH$ radius and mass decrease, and the $BH$ 
temperature increases, as it should be. But here the $BH$ radius is bounded from below (by $r_{\min}$
and the temperature does not blow up (as it is bounded by $T_S$). The ``mass loss'' and ``time life''
are:

$$- \left( \frac{dM}{dt} \right)_+ = - \left( \frac{dM}{dt} \right) \: \left( 1 + \frac{20}{21} \: 
\frac{{\cal A}}{r^6_S} \right)$$

$$\tau_+ = \tau_H \left( 1 - \frac{8}{7} \:  \frac{{\cal A}}{r^6_S} \right) $$

\medskip
\noindent
The life time of the string black hole is $\tau_{\min} = (\frac{K_{BC}}{G \hbar}) T_S^{-3}$.

The string back reaction effect is finite and consistently describes both, the QFT regime ($BH$ of 
mass $M$ and temperature $T_H$) and the string regime ($BH$ of mass $M_{\min}$ and temperature $T_S$).
Both regimes are bounded as in string theory we have:

$$\begin{array}{lll} r_{\min} \leq r_+ \leq r_S \qquad , \qquad M_{\min} \leq M_+ \leq M  
\\
\tau_{\min} \leq \tau_+ \leq \tau_H \qquad , \qquad T_H \leq T_+ \leq T_S  \end{array}$$

\medskip
\noindent
The ${\cal R}$ ``Dual'' transform well summarizes the link between the two opposite well defined
regimes: $T_H \ll T_S$ (ie $r_S \gg r_{\min} , M \gg M_{\min}$) and $T_H \to T_S$ (ie $r_S \to R_{\min}
, M \to M_{\min}$).

This paper is organized as follows: In Section $2$ we summarize the classical $BH$ geometry and its
semiclassical thermal properties in the QFT-Hawking regime. In Section $3$ we derive the bonds imposed 
by string theory on this regime and show the Dual relation between the string and Hawking temperatures. 
In Section $4$ we compute the quantum string emission by the $BH$. In Section $5$ we compute its back
reaction effect. Section $6$ presents conclusions and remarks.

\section{THE SCHWARZSCHILD BLACK HOLE SPACE TIME}

\hspace{0.5cm} The $D -$ dimensional Schwarzschild Black Hole metric reads  

\begin{equation} 
ds^2 = -a(r) c^2 dt^2 + a^{-1} (r) dr^2 + r^2 d \: \Omega^2_{D-2}     
\end{equation}

\noindent
where

\begin{equation}
a(r) = 1 - \left( \frac{r_S}{r} \right)^{D-3}            
\end{equation}

\noindent
being $r_S$ the horizon (or Schwarzschild radius)   

$$r_S = \left( \frac{16 \pi G M}{c^2 (D-2) A_{D-2}} \right)^{\frac{1}{D-3}}   \eqno(3.a) $$

\noindent
and

\setcounter{equation}{3}
\begin{equation}
A_{D-2} = \frac{2 \pi^{\frac{(D-1)}{2}}}{\Gamma \left( \frac{(D-1)}{2} \right)}           
\end{equation}

\noindent
(surface area per unit radius). $G$ is the Newton gravitational constant. For $D = 4$ one has  

$$r_S = \frac{2 G M}{c^2}     \eqno(3.b)  $$

The Schwarzschild Black Hole ($B.H$) is characterized by  its mass $M$
(angular momentum~: $J = 0$~; electric charge~: $Q = 0$). The horizon [Eq. (3)] and the thermodynamical
magnitudes associated to the $B.H$ -- temperature ($T$), entropy ($S$), and specific heat ($C_V$) -- 
are all expressed in terms of $M$ (Table $1$).  

 A brief review of these quantities is the following: in the context of QFT,Black Holes do emit thermal 
radiation at the Hawking temperature given by   

$$T_H = \frac{\hbar \: \kappa}{2 \pi \: k_B \: c}     \eqno(5.a)  $$

\noindent
where

$$\kappa = \frac{c^2 (D-3)}{2 r_S}     \eqno(5.b)  $$

\noindent
is the surface gravity. For $D = 4$,  

$$T_H = \frac{\hbar \: c^3}{8 \pi \: k_B \: G M}      \eqno(5.c)  $$

The $B.H$ Entropy is proportional to the $B.H$ area $A = r^{D-2}_S A_{D-2}$ [Eq. (4)]  

\setcounter{equation}{5}
\begin{equation}
S = \frac{1}{4} \: \frac{k_B \: c^3}{G \: \hbar} A               
\end{equation}

\noindent
and its specific heat $C_V = T \left( \frac{\partial S}{\partial T} \right)_V $ is negative

$$C_V = - \frac{(D-2)}{4} \: \frac{k_B \: c^3}{G \: \hbar} \: A_{D-2} \: r^{D-2}_S   \eqno(7.a)  $$

\noindent
In $4 -$ dimensions it reads  

$$C_V = - \frac{8 \pi \: k_B \: G}{\hbar \: c} \: M^2        \eqno(7.b)   $$

 As it is known, [Eq. (5)], [Eq. (6)] and [Eq. (7)] show that the $B.H -$ 
according to its specific heat being negative -- increases its temperature in its quantum emission 
process ($M$ decreases). Also, it could seem that, if the $B.H$ would evaporate completely ($M = 0$),
the QFT-Hawking temperature $T_H$ would  become infinite.
However, at this limit, and more precisely when $M \sim M_{PL}$, the fixed classical background 
approximation for the $B.H$ geometry breaks down, and the back reaction effect of the radiation matter
on the $B.H$ must be taken into account. In Section $5$, we will take into account this back reaction
effect in the framework of string theory.  

First, we will consider quantum strings in the fixed $B.H$ background. We will see that even in this
approximation, quantum string theory not only
can retard the catastrophic process but, furthermore, provides non-zero lower bounds for the 
$B.H$ mass ($M$) or horizon ($r_S$), and a finite (maximal) value for the $B.H$ temperature $T_H$ as
well.

\section{QUANTUM STRINGS IN THE BLACK HOLE SPACE TIME}

The Schwarzschild black hole spacetime is asymptotically
flat. Black hole evaporation -- and any ``slow down'' of this process
-- will be measured by an  
observer which is at this asymptotic region. In Ref. [2] it has been
found that the mass spectrum of  
quantum string states coincides with the one in Minkowski
space. Critical dimensions are the same as  
well Ref. [2]
($D = 26$, open and closed bosonic strings; $D = 10$ super and
heterotic strings). 

Therefore, the asymptotic string mass density of levels in black hole spacetime
will read as in Minkowski space 
\setcounter{equation}{7}
\begin{equation}
\rho (m) \sim \left( \sqrt{\frac{\alpha^\prime c}{\hbar}} \: m \right)^{-a} \: 
e^{b\sqrt{\frac{\alpha^\prime c}{\hbar}} \: m}                                 
\end{equation}

\noindent
where $\alpha^\prime \equiv \frac{c^2}{2 \pi \: T}$ ($T$~: string
tension) has dimensions of $({\rm  
linear \: mass \: density})^{-1}$~; constants $a / b$ depend on the
dimensions and on the type of  
string Ref. [8]. For a non-compactified space-time these coefficients
are given in Table $2$. 

\subsection{The Maximal Black Hole Temperature}

In this paper, strings in a $B.H$ spacetime are considered in the
framework of the string analogue model. In this model, one considers
the strings as a collection of quantum fields $\phi_1 , \cdots , \phi_n$ , 
whose masses are given by the string mass spectrum ( $\alpha^\prime
(\frac{c}{\hbar}) m^2 \simeq n$ ,  for open strings and large $n$ in
flat spacetime). Each field of mass $m$ appears as many times as the
degeneracy of the mass level~; for higher excited modes this is
described by $\rho (m)$ [Eq. (8)]. Although quantum fields do not interact among themselves, they do with the $B.H$ background.  

In the asymptotic (flat) $B.H$ region, the thermodynamical behavior of the 
higher excited quantum string states of open strings, for example, is
deduced from the canonical partition function Ref. [5]  
\begin{equation}
\ln Z = \frac{V}{(2 \pi )^d} \: \sqrt{\frac{\alpha^\prime c}{\hbar}}
\: \int\limits^{\infty}_{m_0} \: 
dm \: \rho (m) \: \int d^d k \ln \left\lbrace \frac{1 + \exp
\left\lbrack - \beta_H \left( m^2 c^4 + k^2   
\hbar^2 c^2 \right)^{\frac{1}{2}} \right\rbrack}{1 - \exp \left\lbrack
- \beta_H \left( m^2 c^4 + k^2 
\hbar^2 c^2 \right)^{\frac{1}{2}} \right\rbrack} \right\rbrace
\end{equation}
\noindent
($d$: number of spatial dimensions) where supersymmetry has been considered for the sake of generality; $\rho (m)$ 
is the asymptotic mass density given by [Eq. (8)]; $\beta_H = (k_B
T_H)^{-1}$ where $T_H$ is the $B.H$ Hawking temperature; $m_0$ is the
lowest mass for which $\rho (m)$ is valid. 

\vspace{5mm} 

\begin{tabular}{|l|c|c|}
\hline
& & \\
\hspace{2cm} & \hspace{2.5cm} Dimension: $D$ \hspace{2.5cm} & \hspace{2.5cm} $D = 4$ \hspace{2.5cm} \\
& & \\
\hline
& & \\
$r_S$ & $\left\lbrack \frac{16 \pi G M}{c^2 \: (D-2) \: A_{D-2}} \right\rbrack^{\frac{1}{D-3}}$ & 
$\frac{2 G M}{c^2}$ \\
& & \\
\hline
& & \\
$\kappa$ & $\frac{(D-3) \: c^2}{2 r_S}$ & $\frac{c^2}{2 r_S}$ \\
& & \\
\hline
& & \\
$A$ & $A_{D-2} \: r^{D-2}_S$ & $4 \pi \: r^2_S$ \\
& & \\
\hline
& & \\
$T_H$ & $\frac{\hbar \: \kappa}{2\pi \: k_B \: c}$ & $\frac{\hbar \: c}{4\pi \: k_B \: r_S}$ \\
& & \\
\hline
& & \\
$S$ & $\frac{1}{4} \: \frac{k_B \: c^3}{G \: \hbar} \: A$ & $\frac{k_B \: \pi \: c^3}{G \: \hbar} \: 
r^2_S$ \\
& & \\
\hline
& & \\
$C_V$ & $- \frac{(D-2)}{4} \: \frac{k_B \: c^3}{G \: \hbar} \: A_{D-2} \: r^{D-2}_S$ & 
$- \frac{2 \pi \: k_B \: c^3}{G \: \hbar} \: r^2_S$ \\
& & \\
\hline
\end{tabular}

\vspace{5mm} 

\tablename{$\:$ 1}: Schwarzschild black hole thermodynamics. $M$
($B.H$ mass); $r_S$ (Schwarzschild  radius);
$\kappa$ (surface gravity); $A$ (horizon area); $T_H$ (Hawking
temperature); $S$ (entropy); $C_V$ (specific heat); $G$ and $k_B$
(Newton and Boltzman constants); $A_{D-2} = 2 \pi^{\frac{(D-1)}{2}} /
\Gamma \left( \frac{(D-1)}{2} \right)$. 

\vspace{5mm} 

For the higher excited string modes, ie the masses of the $B.H$ and the higher string modes satisfy 
the condition  

$$\beta_H \: m \: c^2 = \frac{4 \pi \: m \: c}{(D-3) \: \hbar} \: \left\lbrack  \frac{16 \pi \: G M}
{c^2 (D - 2) A_{D-2}} \right\rbrack^{\frac{1}{D-3}} \: \gg \: 1
\eqno(10.a)  $$

\noindent
which reads for $D = 4$

$$\beta_H \: m \: c^2 = \frac{8 \pi \: G M \: m}{\hbar \: c} \: \gg \: 1            \eqno(10.b) $$

\noindent
(condition [Eq. (10.b)] will be  considered later in section $4$) the leading contribution to the r.h.s.
of [Eq. (9)] will give as a canonical partition function  

\setcounter{equation}{10}
\begin{equation}
\ln Z \simeq \frac{2 V_{D-1} \: \left( \alpha^\prime \frac{c}{\hbar} \right)^{- \frac{(a-1)}{2}}}
{(2 \pi \: \beta_H \: \hbar^2)^{\frac{D-1}{2}} } \: \cdot \: \int\limits^\infty_{m_0}  d \: m \: 
m^{-a + \frac{D-1}{2}} \: \: e^{-( \beta_H - \beta_S ) m c^2}                  
\end{equation}

\noindent
where $\beta_S = (k_B T_S)^{-1}$ , being $T_S$  [Eq. (8)]

\begin{equation}
T_S = \frac{c^2}{k_B \: b \left( \frac{\alpha^\prime c}{\hbar} \right)^{\frac{1}{2}}}      
\end{equation}


\begin{tabular}{|c|c|c|c|c|}
\hline
& & & & \\
Dimension & String Theory & $a$ & $b$ & $k_B \: T_S / c^2$ \\
& & & & \\
\hline
& & & & \\
$D$ & open bosonic & $(D-1) / 2$ & & \\ \cline{2-3} & closed & $D$ & $2 \pi \sqrt{\frac{D-2}{6}}$ & 
$\left\lbrack 2 \pi \sqrt{\frac{(D-2)}{6} \: \left( \frac{\alpha^\prime c}{\hbar} \right)} 
\right\rbrack^{-1}$ \\
& & & & \\
\hline
& & & & \\
$26$ & & & & \\ (critical) & open bosonic & $25 / 2$ & & \\ \cline{2-3} & closed & $26$ & $4 \pi$ &
$\left( 4 \pi \sqrt{\frac{\alpha^\prime c}{\hbar}} \right)^{-1}$ \\
& & & & \\
\hline
& & & & \\
$10$ & & & & \\ (critical) & open superstring & $9 / 2$ & & \\ \cline{2-3} & closed superstring 
(type II) & $10$   
& $\pi 2 \sqrt{2}$ & $\left( \pi 2 \sqrt{2 \left( \frac{\alpha^\prime c}{\hbar} \right) } \right)^{-1}$ 
\\ \cline{2-3} \cline{4-5}  & Heterotic & $10$ & $\pi (2 + \sqrt{2})$ 
& $\left\lbrack \pi (1 + \sqrt{2}) \sqrt{2 \left( \frac{\alpha^\prime c}{\hbar} \right)} 
\right\rbrack^{-1}$ \\
& & & & \\
\hline
\end{tabular}

\vspace{5mm} 

\tablename{$\:$ 2}: Density of mass levels $\rho (m) \sim m^{-a} \exp
\lbrace b \sqrt{\frac{\alpha^\prime c}{\hbar}} m \rbrace$. For open
strings $\alpha^\prime ( \frac{c}{\hbar} ) m^2 \simeq n$ ; for closed
strings $\alpha^\prime ( \frac{c}{\hbar} ) m^2 \simeq 4n$. 

\vspace{5mm}

\noindent
the string temperature (Table $2$). For open bosonic strings one divides by $2$ the r.h.s. of 
[Eq. (11)] (leading contributions are the same for bosonic and fermionic sector).  

From [Eq. (11)] we see that the definition of $\ln Z$ implies the following condition on the Hawking 
temperature  

\begin{equation}
T_H < T_S              
\end{equation}

 Furthermore, as $T_H$ depends on the $B.H$ mass $M$, or on the horizon $r_S$~, 
[Eq. (5.a)], [Eq. (5.b)] and [Eq. (3)], the above condition will lead to further conditions on the 
horizon. Then $T_S$ represents a critical value temperature: $T_S \equiv T_{cr}$. In order to see this
more clearly, we rewrite $T_S$ in terms of the quantum string length scale  

\begin{equation}
L_S = \left( \frac{\hbar \: \alpha^\prime}{c} \right)^{\frac{1}{2}}          
\end{equation}

\noindent
namely

\begin{equation}
T_S = \frac{\hbar \: c}{b \: k_B \: L_S}            
\end{equation}

 From [Eq. (13)], and with the help of [Eq. (5.a)], [Eq. (5.b)] and [Eq. (15)], we 
deduce  

\begin{equation}
r_S > \frac{b \: (D - 3)}{4 \pi} \: L_S          
\end{equation}

\noindent
which shows that (first quantized) string theory provides a lower bound, or {\bf minimum radius}, 
for the $B.H$ horizon.

 Taking into account [Eq. (3)] and [Eq. (16)] we have the following condition on the
$B.H$ mass  

\begin{equation}
M > \frac{c^2 \: (D - 2) A_{D-2}}{16 \pi \: G} \: \left\lbrack \frac{b
\: (D - 3)}{4 \pi} \: L_S  
\right\rbrack^{D-3}                       
\end{equation}

\noindent
therefore there is a {\bf minimal $B.H$ mass} given by string theory.

\hspace{0.5cm} For $D = 4$ we have 

\begin{eqnarray}
r_S &>& \frac{b}{4 \pi} \: L_S            \\
M &>& \frac{c^2 \: b}{8 \pi \: G} \: L_S           
\end{eqnarray}

\noindent
these lower bounds satisfy obviously [Eq. (3.b)]. [Eq. (19)] can be
rewritten as

$$M > \frac{b \: M^2_{PL}}{8 \pi \: M_S} $$

\medskip
\noindent
where $M_S = \frac{\hbar}{L_S c} $ is the string mass scale ($L_S$~: reduced Compton wavelenght)
and $M_{PL} \equiv \left( \frac{\hbar c}{G} \right)^{\frac{1}{2}} $ is the Planck mass. The minimal
$B.H$ mass is then [Eq. (14)] and [Eq. (19)]  

$$ M_{\min} = \frac{b}{8 \pi G} \: \sqrt{\hbar c^3 \alpha^{\prime}} $$

It is appropriate, at this point, to make use of the ${\cal R}$ or Dual transformation over a length 
introduced in Ref. [3]. This operation is

$$
\tilde{L}_{cl} = {\cal R} L_{cl} = {\cal L}_{\cal R} L^{-1}_{cl} = L_q
\quad {\rm and} \quad
\tilde{L}_q = {\cal R} L_q = {\cal L}_{\cal R} L^{-1}_q = L_{cl}  \eqno(20.a)
$$

\medskip
\noindent
where ${\cal L}_{\cal R}$ has dimensions of $({\rm lenght})^2$; and it is given by ${\cal L}_{\cal R}
= L_{cl} L_q$.

In our case, $L_{cl}$ is the classical Schwarzschild radius, and $L_Q \equiv r_{\min} = (b (D-3) L_S)
/ 4 \pi$ [Eq. (16)].  

The ${\cal R}$ transformation links classical lengths to quantum string lenghts, and more generally it 
links QFT and string theory domains Ref. [3].  

and the string temperature is

$$T_H = \: \frac{\hbar c (D - 3)}{4 \pi k_B L_{cl}}         \eqno(20.b) $$

\medskip
\noindent
For the $BH$, the QFT-Hawking temperature is

$$T_S = \: \frac{\hbar c (D - 3)}{4 \pi k_B L_q}         \eqno(20.c) $$

\medskip
\noindent
Under the ${\cal R}$ operation we have

$$
 \tilde{T}_H = T_S \qquad {\rm and}\qquad
\tilde{T}_S = T_H \eqno(20.d)  
$$

\noindent
which are valid for all $D$.

From the above equations we can read as well

$$\tilde{T}_H \: \tilde{T}_S = T_S \: T_H  $$

We see that under the ${\cal R}$-Dual operation, the QFT temperature and the string temperature in the 
$BH$ background are mapped one into another. This appears to be a general feature for QFT and string 
theory in curved backgrounds, as we have already shown this relation in the de Sitter background
Ref. [3].  

It is interesting to express $T_H$ and $T_S$ in terms of their respective masses

$$\begin{array}{ccc}
T_H & = & \: \frac{\hbar c (D - 3)}{4 \pi k_B} \: \left( \frac{16 \pi G M}{c^2 (D - 2) A_{D-2}}
\right)^{- \frac{1}{D-3}}   \\
& & \\
T_H & = & \: \frac{\hbar c^3}{8 \pi k_B G M}     \qquad \qquad (D = 4) \end{array}$$

and

$$ T_S = \: \frac{c^2 M_S}{b k_B}   $$

\section{QUANTUM STRING EMISSION OF THE BLACK HOLE}

\hspace{0.5cm} As it is known, thermal emission of massless particles
by a black hole has been  
considered in the context of QFT Ref. [1], Ref. [9], Ref. [10].
Here, we are going to deal with thermal emission of high massive
particles which correspond to the 
higher excited modes of a string. The study will be done in the
framework of the string analogue model. 

For a static $D -$ dimensional black hole, the quantum emission cross section $\sigma_q (k,D)$
is related to the total classical absorption cross section $\sigma_A (k,D)$ through the Hawking 
formula Ref. [1]  

\setcounter{equation}{20}
\begin{equation}
\sigma_q (k,D) = \frac{\sigma_A (k,D)}{e^{E(k) \beta_H} - 1}        
\end{equation}

\noindent
where $E(k)$ is the energy of the particle (of momentum~: $p = \hbar \: k$) and $\beta_H = 
(k_B T_H)^{-1}$,
being $T_H$ Hawking temperature [Eq. (5.a)] and [Eq. (5.b)].

The total absorption cross section $\sigma_A (k,D)$ in [Eq. (21)] has two terms Ref. [6],
one is an isotropic $k -$ independent part, and the other has an oscillatory behavior, as a function
of $k$, around the optical geometric constant value with decreasing amplitude and constant period. 
Here we will consider only the isotropic term, which is the more relevant in our case.  

\noindent
For a $D -$ dimensional black hole space-time, this is given by (see for example Ref.[2])  

\begin{equation}
\sigma_A (k,D) = a(D) \: r^{D-2}_S            
\end{equation}

\noindent
where $r_S$ is the horizon [Eq. (3.a)] and [Eq. (4)] and  

\begin{equation}
a(D) = \frac{\pi^{\frac{(D-2)}{2}}}{\Gamma \left( \frac{D-2}{2} \right)} \: \left( \frac{D-1}{D-3} 
\right) \: \left( \frac{D-1}{2} \right)^{\frac{2}{D-3}}         
\end{equation}

We notice that $\rho (m)$ [Eq. (8)] depends only on the mass, therefore we could consider, in our 
formalism, the emitted high mass spectrum as spinless. On the other hand, as we are dealing with
a Schwarzschild black hole (angular momentum equal to zero), spin considerations can be overlooked. 
Emission is larger for spinless particles Ref. [11].

The number of scalar field particles of mass $m$ emitted per unit time is  

\begin{equation}
\langle n(m) \rangle = \int^\infty_0 \langle \: n(k) \rangle \: d \mu (k)   
\end{equation}

\noindent
where $d \mu (k)$ is the number of states between $k$ and $k + dk$.

\begin{equation}
d \mu (k) = \frac{V_d}{(2 \pi)^d} \: \frac{2 \pi^{\frac{d}{2}}}{\Gamma \left( \frac{d}{2} \right)} \:
k^{d-1} \: dk      
\end{equation}

\noindent
and $\langle n(k) \rangle$ is now related 
to the quantum cross section $\sigma_q$ ([Eq. (21)] and [Eq. (22)]) through the equation.  

\begin{equation}
\langle n(k) \rangle = \frac{\sigma_q (k,D)}{r^{D-2}_S}       
\end{equation}

Considering the isotropic term for $\sigma_q$ [Eq. (22)] and [Eq. (23)] we have

\begin{equation}
\langle n(k) \rangle = \frac{a(D)}{e^{E(k) \beta_H} - 1}       
\end{equation}

\noindent
where $\beta_H = \frac{1}{k_B T_H}$, being $T_H$ the ${\rm BH}$ temperature [Eq. (5)].

From [Eq. (24)] and [Eq. (27)], $\langle n(m) \rangle$ will be given by  

\begin{equation}
\langle n(m) \rangle = F(D,\beta_H) \: m^{\frac{(D-3)}{2}} \: (m c^2 \beta_H + 1) e^{- \beta_H
 m c^2}                                                        
\end{equation}

\noindent
where  

\begin{equation}
F(D,\beta_H) \equiv \: \frac{V_{D-1} a(D)}{(2 \pi)^{\frac{(D-1)}{2}}} \: \frac{(c^2)^{\frac
{(D-3)}{2}}}{\beta^{\frac{(D+1)}{2}}_H \: (\hbar c)^{(D-1)}}   \: \equiv A(D) 
\beta_H^{- \frac{D+1}{2}}                                                                  
\end{equation}

\noindent
Large argument $\beta_H \: m \: c^2 \gg 1$, ie [Eq. (10.a)] and [Eq. (10.b)], 
and leading approximation have been considered in performing the $k$ integral.

The quantum thermal emission cross section for particles of mass $m$ is defined as

$$\sigma_q (m,D) = \int \sigma_q (k,D) d \mu (k)         \eqno(30.a) $$

\noindent
and with the help of [Eq. (26)] we have  

$$\sigma_q (m,D) = r^{D-2}_S \: \langle n(m) \rangle       \eqno(30.b) $$

\noindent
where $\langle n(m) \rangle$ is given by [Eq. (28)].

In the string analogue model, the string quantum thermal emission by a ${\rm BH}$ will be given by the
cross section  

\setcounter{equation}{30}
\begin {equation}
\sigma_{\rm string} (D) = \: \sqrt{\frac{\alpha^\prime c}{\hbar}} \: \int^\infty_{m_0} \: 
\sigma_q (m,D) \: \rho (m) dm       
\end{equation}

\noindent
where $\rho (m)$ is given by [Eq. (8)], and $\sigma_q (m,D)$ by [Eq. (30)] and [Eq. (28)];
$m_0$ is the lowest string field mass for which the asymptotic value of the density of mass levels, 
$\rho (m)$, is valid.

For arbitrary $D$ and $a$, we have from [Eq. (31)], [Eq. (30.b)], [Eq. (28)] and [Eq. (8)]  

\begin{equation}
\sigma_{\rm string} (D) = F (D,\beta_H) r^{D-2}_S \: \left( \sqrt{\frac{\alpha^\prime c}{\hbar}}
\right)^{-a+1} \: I_D (m,\beta_H - \beta_{cr}, a)              
\end{equation}

\noindent
where $F (D, \beta_H)$ is given by [Eq. (29)] and  

$$\beta_{cr} \equiv \beta_S = (k_B T_S)^{-1} = \frac{b}{c^2} \: \sqrt{\frac{\alpha^{\prime} c}{\hbar}}
\eqno(33.a) $$

and  

$$I_D (m,\beta_H - \beta_{cr}, a) \equiv \: \int^\infty_{m_0} \: m^{-a + \frac{D-3}{2}} \: (m c^2 
\beta_H + 1) \: e^{-(\beta_H - \beta_{cr}) m c^2} \: dm                 \eqno(33.b)   $$

After a straightforward calculation we have 

$$
I_D (m,\beta_H - \beta_S, a)  = \: \frac{c^2 \beta_H}{\lbrack
(\beta_H - \beta_S)  
c^2 \rbrack^{-a + \frac{(D+1)}{2}}} \: \Gamma \: \left( -a +
\frac{D+1}{2}, (\beta_H - \beta_S) 
c^2 m_0 \right) 
$$
$$
+ \: \frac{1}{\lbrack (\beta_H - \beta_S) c^2 \rbrack^{-a + \frac{(D-1)}
{2}}} \: \Gamma \: \left( -a + \frac{D-1}{2}, (\beta_H - \beta_S) c^2
m_0 \right)    
 \eqno(33.c)   
$$ 

\noindent
where $\Gamma (x,y)$ is the incomplete gamma function.

For open strings, $a = \frac{(D-1)}{2}$ [$D$~: non-compact dimensions], we have  

\setcounter{equation}{33}
\begin{equation}
\sigma^{({\rm open})}_{\rm string} (D) = A(D) \beta_H^{- \frac{(D+1)}{2}}  \: r^{D-2}_S \: \left( 
\frac{c^2 \beta_S}{b} \right)^{-\frac{(D-3)}{2}} \: \cdot \:
\lbrace \frac{\beta_H}{\beta_H - \beta_S} \: e^{- (\beta_H - \beta_S) c^2 m_0} \: - \: E_i \left(
- (\beta_H - \beta_S) c^2 m_0 \right) \rbrace             
\end{equation}

\noindent
where $E_i$ is the exponential-integral function, and we have used [Eq. (29)] and [Eq. (33.A)].

When $T_H$ approaches the limiting value $T_S$, and as $E_i (-x) \sim C + \ln x$ for small $x$,
we have from [Eq. (34)]

$$\begin{array}{lcl}
\sigma^{({\rm open})}_{\rm string} (D) & = & A(D) \beta_S^{- \frac{(D+1)}{2}}  \: r^{D-2}_{\min} \: 
\left( \frac{c^2}{b} \beta_S \right)^{- \frac{(D-3)}{2}} \\
& & \\
&  & \: \lbrace \frac{\beta_S}{\beta_H - \beta_S} \: - C - \ln \: \left( (\beta_H - \beta_S) m_0
c^2 \right) \rbrace  \\
& & \\
& = & B (D) \beta^{-1}_S \: \lbrace \frac{\beta_S}{\beta_H - \beta_S} \: - C - \ln \: \left( (\beta_H - 
\beta_S) c^2 m_0 \right) \rbrace  \end{array} $$

\medskip
\noindent
where \\

$$r_{\min} = \: \frac{\hbar c (D-3) \beta_S}{4 \pi}$$

\medskip
\noindent
and  

$$B(D) \equiv A(D) \left( \frac{\hbar c (D-3)}{4 \pi} \right)^{D-2} \: \left( \frac{c^2}{b}
\right)^{- \frac{(D-3)}{2}}                         \eqno(34.a)  $$

For $\beta_H \to \beta_S$ the dominant term is 
$$
\sigma^{\rm (open)}_{\rm string} (D) \:  \buildrel{T_H \longrightarrow
T_S}\over\simeq   B(D) \:\frac{1}{(\beta_H - \beta_S)}  \eqno(35.a) 
$$

\medskip
\noindent
for any dimension.

For $\beta_H \gg \beta_S$, ie $T_H \ll T_S$

$$
\sigma^{({\rm open})}_{\rm string} (D)  \buildrel{T_H  \ll
T_S}\over\simeq\simeq  \: A(D) \beta_H^{-
\frac{(D+1)}{2}}  \: r^{D-2}_S \:  
\left( \frac{c^2}{b} \beta_S \right)^{-\frac{(D-3)}{2}} 
\: e^{- \beta_H c^2 m_0} \: \left( 1 + \: \frac{1}{\beta_H c^2 m_0} \right)
$$
$$
 \simeq  \: B(D) \beta_H^{\frac{(D-5)}{2}} \: \beta_S^{-
\frac{(D-3)}{2}} \: e^{- \beta_H c^2 m_0} \eqno(35.b)
$$

\medskip
\noindent
as $E_i (-x) \sim \: {e^{-x}\over x} + \cdots $ for large $x$. For $D = 4$,

$$\sigma^{({\rm open})}_{\rm string} (4)  \simeq  \: B(4) \left( \frac{1}{\beta_H \beta_S} 
\right)^{\frac{1}{2}} \: e^{- \beta_H c^2 m_0} $$

At this point, and in order to interpret the two different behaviours, we compare them with the 
corresponding behaviours for the partition function [Eq. (11)]. For open strings ($a = (D-1)/2$)
$\ln Z$ is equal to

$$\ln \: {\cal Z}_{\rm open} \: \simeq \: \frac{2 V_{D-1} \left(
\frac{\alpha^{\prime} c}{\hbar}  
\right)^{- \frac{(D-3)}{4}}}{(2 \pi \beta_H \hbar^2)^{(D-1)\over 2}} \:
\cdot \: \frac{1}{(\beta_H -  
\beta_S) c^2} \: \cdot \: e^{-(\beta_H - \beta_S ) m_0 c^2} $$

For $\beta_H \to \beta_S$:

$$\begin{array}{ccc}
\ln \: {\cal Z}_{\rm open} & \: \simeq \: & \frac{2 V_{D-1} \left( \frac{\alpha^{\prime} c}{\hbar} 
\right)^{- \frac{(D-3)}{4}}}{(2 \pi \beta_H \hbar^2)^{(D-1) \over 2}}
\: \cdot \: \frac{1}{(\beta_H -  \beta_S) c^2}   \\
& \beta_H \to \beta_S &  \end{array} $$

\noindent
and for $\beta_H \gg \beta_S$:

$$\begin{array}{ccc}
\ln \: {\cal Z}_{\rm open} & \: \simeq \: & \frac{2 V_{D-1} \left(
\frac{\alpha^{\prime} c}{\hbar}\right)^{- \frac{(D-3)}{4}}}{{(2 \pi
\hbar^2)^{(D-1) \over 2}} c^2 
\beta_H^{(D+1) \over 2}} \: \cdot \:  e^{- \beta_H  m_0 c^2} \\
& \beta_H \gg \beta_S &  \end{array} $$

The singular behaviour for $\beta_H \to \beta_S$, and all $D$, is typical of a string system with
intrinsic Hagedorn temperature, and indicates a string phase transition (at $T = T_S$) to a condensed
finite energy state (Ref. [5]). This would be the minimal black hole, of mass $M_{\min}$ and
temperature $T_S$.

\section{QUANTUM STRING BACK REACTION}

\hspace{0.5cm} When we consider quantised matter on a classical background, the dynamics can be 
described by the following Einstein equations

\begin{equation}
R^\nu_\mu - \frac{1}{2} \: \delta^\nu_\mu R = \frac{8 \pi G}{c^4} \: \langle \tau^\nu_\mu \rangle 
\end{equation}

\noindent
The space-time metric $g_{\mu\nu}$ generates a non-zero vacuum expectation value of the energy momentum
tensor $\langle \tau^\nu_\mu \rangle$, which in turn, acting as a source, modifies the former 
background. This is the so-called back reaction problem, which is a semiclassical approach to the 
interaction between gravity and matter.   

\medskip
Our aim here is to study the back reaction effect of higher massive (open) string modes (described
by $\rho (m)$ , [Eq. (8)]) in black holes space-times. This will give us an insight on the last stage
of black hole evaporation. Back reaction effects of massless quantum fields in these equations were 
already investigated Ref. [12], Ref. [13], Ref. [14].

As we are also interested in stablishing the differences, and partial analogies, between string theory
and the usual quantum field theory for the back reaction effects in black holes space-times, we will 
consider a $4 -$ dimensional physical black hole.  

\medskip
The question now is how to write the appropriate energy-momentum tensor $\langle \tau^\nu_\mu \rangle$
for these higher excited string modes. For this purpose, we will consider the framework of the string
analogue model. 
In the spirit of this model, the v.e.v. of the stress tensor $\langle \tau^\nu_\mu \rangle$ for the 
string higher excited modes is defined by  

\begin{equation}
\langle \tau^\nu_\mu (r) \rangle = \frac{\int^\infty_{m_0} \: \langle T^\nu_\mu (r,m) \rangle \:
\langle n(m) \rangle \rho (m) \: dm}{\int^\infty_{m_0} \: \langle n(m) \rangle \rho (m) \: dm}   
\end{equation}

\noindent
where $\langle T^\nu_\mu (r,m) \rangle$ is the Hartle-Hawking vacuum expectation value of the stress 
tensor of an individual quantum field, and the $r$ dependence of $\langle T^\nu_\mu \rangle$ preserves
the central gravitational character of the problem~; $\rho (m)$ is the string mass density of levels
[Eq. (8)] and $\langle n(m) \rangle$ is the number of field particles of mass $m$ emitted per unit 
time, [Eq. (28)].

\hspace{0.5cm} In order to write $\langle T^\mu_\mu (m,r) \rangle$ for an individual quantum field in 
the framework of the analogue model, we notice that $\rho (m)$ [Eq. (8)] depends only on $m$~; 
therefore, we will consider for simplicity the vacuum expectation value of the stress tensor for a 
massive scalar field.  

\medskip
For the Hartle-Hawking vacuum (black-body radiation at infinity in equilibrium with a black 
hole at the temperature $T_H$), and when the (reduced) Compton wave length of the massive particle 
$(\lambda = \frac{\hbar}{mc})$ is much smaller than the Schwarzschild radius $(r_S)$  

\setcounter{equation}{41}
\begin{equation}
\frac{\hbar \: c}{2 G M m} \ll 1        
\end{equation}

\noindent
(same condition as the one of [Eq. (10.b)]), $\langle T^r_r \rangle$ and $\langle T^0_0 \rangle$ for 
the background $B.H$ metric ([Eq. (1)], $D = 4$) read Ref. [13]  

$$\frac{8 \pi G}{c^4} \: \langle T^r_r \rangle = \frac{A}{r^8} \: F_1 \left( \frac{r_S}{r} \right)
\eqno(43.a)  $$

$$\frac{8 \pi G}{c^4} \: \langle T^0_0 \rangle = \frac{A}{r^8} \: F_2 \left( \frac{r_S}{r} \right)
\eqno(43.b)  $$

\noindent
where 

$$A = \frac{M^2 \: L^6_{PL}}{1260 \pi m^2}        \eqno(43.c)  $$

$$F_1 \left( \frac{r_S}{r} \right) = 441 - \zeta 2016 + \frac{r_S}{r} \: \left( -329 + \zeta 1512 
\right) + O (m^{-4})               \eqno(43.d)  $$

$$F_2 \left( \frac{r_S}{r} \right) = -1125 + \zeta 5040 + \frac{r_S}{r} \left( 1237 - \zeta 5544 
\right) + O (m^{-4})                 \eqno(43.e)  $$

\noindent
$M$ and $m$ are the black hole and the scalar field masses respectively, $\zeta$ (a numerical factor)
is the scalar coupling parameter ($- \frac{\zeta R \phi^2}{2}$~; $R$ : scalar curvature, $\phi$ : 
scalar field) and $L_p \equiv (\frac{\hbar G}{c^3})^{\frac{1}{2}}$ is the Planck length.

From [Eq. (36)], [Eq. (43.a)] and [Eq. (43.b)] the v.e.v. of the string stress tensor will read  

$$\frac{8 \pi G}{c^4} \: \langle \tau^r_r \rangle = \frac{\cal A}{r^8} \: F_1  \left( \frac{r_S}{r} 
\right)
\eqno(44.a) $$

$$\frac{8 \pi G}{c^4} \: \langle \tau^0_0 \rangle = \frac{\cal A}{r^8} \: F_2  \left( \frac{r_S}{r} 
\right)
\eqno(44.b) $$

\noindent
where 

\setcounter{equation}{44}
\begin{equation}
{\cal A} = \frac{M^2 \: L^6_{PL}}{1260 \pi} \: \frac{\int^\infty_{m_0} \: m^{-2} \: \langle n(m) \rangle \:
\rho (m) \: dm}{\int^\infty_{m_0} \: \langle n(m) \rangle \: \rho (m) \: dm}        
\end{equation}

We obtain [15] 

\begin{equation}
g^{-1}_{rr} = 1 - \frac{r_S}{r} \: - \: \frac{\cal A}{21 r^6} \: \lbrack 23 \left( \frac{r_S}{r} 
\right) - 27 \rbrack                                 
\end{equation}

From the above equation it is clear that the quantum matter back reaction modifies the horizon, $r_+$,
which will be no longer equal to the classical Schwarzschild radius $r_S$. The new horizon will
satisfy  

$$g^{-1}_{rr} \: = \: 0                   \eqno(51.a)    $$

\noindent
ie 

$$r^7_+ - r_S r^6_+ + {\cal A} \: \frac{27}{21} \: r_+ - {\cal A} \: \frac{23}{21} \: r_S = 0       
\eqno(51.b)  $$

In the approximation we are dealing with ($O (m^{-4})$ ie ${\cal A}^2 \ll {\cal A}$), the solution 
will have the form 

\setcounter{equation}{51}
\begin{equation}
r_+ \simeq r_S ( 1 + \epsilon )  \: , \: \epsilon \ll 1                  
\end{equation}

\noindent
From [Eq. (51.b)] we obtain  

\begin{equation}
r_+ \simeq r_S \left( 1 - \: \frac{4 {\cal A}}{21 r^6_S} \right)           
\end{equation}

\noindent
which shows that the horizon decreases.

Let us consider now the surface gravity, which is defined as  

\begin{equation}
k (r_+) = \: \frac{c^2}{2} \: \frac{d g^{-1}_{rr}}{dr} \vert_{r=r_+}           
\end{equation}

\noindent
(in the absence of back reaction, $k (r_+) = k (r_S)$ given by [Eq. (5.b)] for $D = 4$).

From [Eq. (50)], [Eq. (53)] and [Eq. (54)] we get  

\begin{equation}
k (r_+) = \: \frac{c^2}{2 r_S} \: \left( 1 + \frac{1}{3} \: \frac{\cal A}{r^6_S} \right)      
\end{equation}

The black hole temperature will then be given by  

\begin{equation}
T_+ = \: \frac{\hbar \kappa (r_+)}{2 \pi k_B c} \: \simeq T_H \: \left( 1 + \frac{1}{3} \: 
\frac{\cal A}{r^6_S} \right)                                                   
\end{equation}

\noindent
where $\: T_H = \: \frac{\hbar c}{4 \pi k_B r_S}$

\noindent
([Eq. (5.a)] and [Eq. (5.b)] for $D = 4$).The Black hole temperature increases due to the back 
reaction.

Due to the quantum emission the black hole suffers a loss of mass. The mass loss rate is given by a
Stefan-Boltzman relation. Without back reaction, we have

\begin{equation}
- \left( \frac{dM}{dt} \right) = \sigma 4 \pi r^2_S T^4_H                    
\end{equation}

\noindent
where $\sigma$ is a constant.

When back reaction is considered, we will have  

\begin{equation}
- \left( \frac{dM}{dt} \right)_+ = \sigma 4 \pi r^2_+ T^4_+             
\end{equation}

\noindent
where $r_+$ is given by [Eq. (53)] and $T_+$ by [Eq. (56)]. Inserting these values into the
above equation we obtain  

\begin{equation}
- \left( \frac{dM}{dt} \right)_+ \simeq - \left( \frac{dM}{dt} \right) \: \left( 1 + \frac{20 {\cal A}}
{21 r^6_S} \right)                                         
\end{equation}

On the other hand, the modified black hole mass is given by  

\begin{equation}
M_+ \equiv \: \frac{c^2}{2 G} \: r_+ \simeq M \left( 1 - \frac{4 {\cal A}}{21 r^6_S} \right)    
\end{equation}

\noindent
which shows that the mass decreases.

From [Eq. (59)] and [Eq. (60)], we calculate the modified life time of the black hole due to the back 
reaction  

\begin{equation}
\tau_+ \simeq \tau_H \left( 1 - \frac{8 {\cal A}}{7 r^6_S} \right)             
\end{equation}

\noindent
We see that $\tau_+ < \tau_H $ since ${\cal A} > 0$.

The string back reaction `form factor' ${\cal A}$ [Eq. (45)] 
can be rewritten  as  

\begin{equation}
{\cal A} = \: \frac{M^2 L^6_{PL}}{1260 \pi} \: \cdot \: \frac{N}{De}                 
\end{equation}

\noindent
where  

\begin{equation}
N = \int^\infty_{m_0} \: m^{-a + \frac{D-7}{2}} \: (m c^2 \beta_H + 1) \: e^{- ( \beta_H - \beta_S)
m c^2} \: dm                                                                 
\end{equation}

\noindent
and [Eq. (33)]  

\begin{equation}
De = I_D (m, \beta_H - \beta_S , a)                 
\end{equation}

For $\beta_H \to \beta_S$ ($M \to M_{\min} , r_S \to r_{\min}$)
we have for open strings  

\begin{equation}
{\cal A}_{\rm open} \simeq \: \frac{M^2_{\min} L^6_{PL} ( \beta_H - \beta_S)}{1260 \pi \: \beta_S} \: 
\left( \frac{1}{2 m^2_0} \: + \: \frac{c^2 \beta_S}{m_0} \right)                               
\end{equation}

Although the string analogue model is in the spirit of the canonical ensemble-all (higher) massive
string fields are treated equally -- we will consider too, for the sake of completeness, the string 
``form factor'' ${\cal A}$ for closed strings.  

\medskip
For $a = D$ ($D$: non compact dimensions), from [Eq. (33.b)],
[Eq. (33.c)] and [Eq. (64)], we have the expressions  for 
$ De = I_D (m, \beta_H - \beta_S , D)$. Its explicit form is given in
ref. [15].

\noindent

For closed strings for $D = 4$, and for $\beta_H \to \beta_S $
($M \to M_{\min} , r_S \to r_{\min}$), we have

\begin{equation}
{\cal A}_{\rm closed} \: = \: \frac{M^2_{\min} L^6_{PL}}{1260 \pi m^2_0} \: \left( \frac{\frac{c^2 
\beta_S}{7} \: + \: \frac{1}{9 m_0}}{\frac{c^2 \beta_S}{3} \: + \: \frac{1}{5 m_0}} \right)     
\end{equation}

From [Eq. (62)] and [Eq. (63)], we evaluate now the number ${\cal A} /
r^6_S$ appearing in the  
expressions for $r_+$ [Eq. (53)], $T_+$ [Eq. (56)], $M_+$ [Eq. (60)], and $\tau_+$ [Eq. (61)], for the
two opposite limiting regimes $\beta_H \to \beta_S$ and $\beta_H \gg \beta_S$:

\begin{eqnarray}
\left( \frac{{\cal A}_{\rm open}}{r^6_{\min}} \right)_{\beta_H \to
\beta_S}  \: & \simeq \: &  
\frac{(\beta_H - 
\beta_S )}{\beta_S} \: \cdot \: \frac{16}{315} \: \left( \frac{\pi}{b} \right)^3 \: \left( \frac{M_S}
{M_{PL}} \right)^2 \: \left( \frac{M_S}{m_0} \right)  \nonumber \\
& \simeq \: & (\beta_H - \beta_S ) \: \cdot \: \frac{16}{315} \: \left( \frac{\pi}{b} \right)^3 \: 
\left( \frac{M_S}{M_{PL}} \right)^2 \:  \frac{M_S^2 c^2}{b m_0}  \: \ll \: 1              
\end{eqnarray}

\begin{equation}
\left( \frac{{\cal A}_{\rm closed}}{r^6_{\min}} \right)_{\beta_H \to \beta_S}  \:  \simeq \: 
\frac{16}{735 b} \: \left( \frac{\pi}{b} \right)^3 \: \left( \frac{M_S}{M_{PL}} \right)^2 \: \left( 
\frac{M_S}{m_0} \right)^2 \: \ll \: 1                                                    
\end{equation}

In the opposite (semiclassical) regime $\beta_H \gg \beta_S$ i.e $T_H
\ll T_S$, we have [15]: 

\begin{equation}
\left( \frac{N}{De} \right)^{\rm open}_{\beta_H \gg \beta_S} \: \simeq \frac{1}{m^2_0} \: \simeq \:
\left( \frac{N}{De} \right)^{\rm closed}_{\beta_H \gg \beta_S}                             
\end{equation}

\noindent
as

\begin{equation}
\beta_H m_0 c^2 = 8 \pi \left( \frac{M}{M_{PL}} \right) \: \left( \frac{m_0}{M_{PL}} \right) \: \gg \:
1                                                                      
\end{equation}

\noindent
$(m_0 , M \gg M_{PL} )$. Then, from [Eq. (62)]

\begin{equation}
\left( \frac{{\cal A}}{r^6_S} \right)^{\rm open / closed}_{\beta_H \gg \beta_S} \:  \simeq \: 
\frac{1}{80640 \pi} \: \left( \frac{M_{PL}}{M} \right)^4 \: \left( \frac{M_{PL}}{m_0} \right)^2 \: 
\ll 1                                                                 
\end{equation}

\noindent
That is, in this regime, we consistently recover $r_+ \simeq r_S$, $T_+ \simeq T_H$, $M_+ \simeq M$
and $\tau_+ \simeq \tau_H = \: \frac{k_B c}{6 \sigma G \hbar T^3_H}$ .

\section{CONCLUSIONS}

We have suitably combined QFT and quantum string theory in the black hole background in the framework of 
the string analogue model (or thermodynamical approach).  

We have computed the quantum string emission by a black hole and the back reaction effect on the black 
hole in the framework of this model. A clear and precise picture of the black hole evaporation emerges.

The QFT semiclassical regime and the quantum string regime of black holes have been identified and
described.  

The Hawking temperature $T_H$ is the intrinsic black hole temperature in the QFT semiclassical regime.
The intrinsic string temperature $T_S$ is the black hole temperature in the quantum string regime. The
two regimes are mapped one into another by the ${\cal R}$ - ``Dual'' transform.  

String theory properly describes black hole evaporation: because of the emission, the semiclassical 
$BH$ becomes a string state (the ``minimal'' $BH$), and the emitted string gas becomes a condensed 
microscopic state (the ``minimal'' $BH$) due to a phase transition. The last stage of the radiation 
in string theory, makes such a transition possible.  

The phase transition undergone by the string gas at the critical temperature $T_S$ represents (in the 
thermodynamical framework) the back reaction effect of the string emission on the $BH$.  

The ${\cal R}$ - ``Dual'' relationship between QFT black holes and quantum strings revealed itself
very interesting. It appears here this should be promoted to a Dynamical operation: evolution from
classical to quantum (and conversely).  

Cosmological evolution goes from a quantum string phase to a QFT and
classical phase. Black hole 
evaporation goes from a QFT semiclassical phase to a string phase. The
Hawking temperature, which we 
know as the black hole temperature, becomes the string temperature for
the `string black hole' in the quantum (string) regime. 

\bigskip

{\Large \bf References}

\bigskip

1. S. W. Hawking, Comm. Math. Phys. 43 (1975) 199.
\medskip

2. H.J. de Vega, N. S\'anchez, Nucl. Phys. B309 (1988) 522; B309 (1988) 577.
\medskip

3. M. Ramon Medrano and N. S\'anchez, 
Phys.Rev. D60 (1999) 125014. 
\medskip

4. R. Hagedorn, Suppl. Nuovo Cimento 3, 147 (1965).
\medskip

5. R.D. Carlitz, Phys. Rev. 5D (1972) 3231.
\medskip

6. N. S\'anchez, Phys. Rev. D18 (1978) 1030.
\medskip

7. M.J. Bowick, L.S. Smolin, L.C. Wijewardhana, Phys. Rev. Lett. 56,
   424 (1986). 
\medskip

8. K. Huang, S. Weinberg, Phys. Rev. Lett 25, 895 (1970).

M.B. Green, J.H. Schwarz, E. Witten, ``{\it Superstring Theory}'', Vol
I. \\ Cambridge University Press, 1987.
\medskip

9. G.W. Gibbons in ``{\it General Relativity, An Einstein Centenary
Survey}'', \\ Eds. S.W. Hawking and  
W. Israel, Cambridge University Press, UK (1979).
\medskip

10. N.D. Birrell, P C W Davies, ``{\it Quantum Fields in Curved
Space}'', \\ Cambridge University Press, UK (1982).
\medskip

11. D.N. Page, Phys. Rev. D13 (1976) 198.
\medskip

12. J.M. Bardeen, Phys. Rev. Lett. 46 (1981) 382.
\medskip

13. V.P. Frolov and A.I. Zelnikov Phys. Lett B115 (1982) 372.
\medskip

14. C.O. Lousto and N. S\'anchez, Phys. Lett B212 (1988) 411 and\\
    Int. J. Mod. Phys. A4 (1989) 2317. 

15. M. Ramon Medrano and N. S\'anchez, Phys. Rev. D61, 084030 (2000).

\end{document}